\documentclass[slac_one]{revtex4}
\usepackage{graphicx}
\usepackage{fancyhdr}
\def\eq#1{Eq.~(\ref{#1})}
\def\beq{\begin{equation}}
\def\eeq{\end{equation}}
\def\beqa{\begin{eqnarray}}
\def\eeqa{\end{eqnarray}}

\newcommand{\as}{\alpha_s}

\newcommand{\e}{\epsilon}

\def \MS {\overline{MS}}

\begin{document}


\title{Angularities and other shapes}

%

\author{Lorenzo Magnea}
\affiliation{Dipartimento di Fisica Teorica, Universit\`a di Torino
and INFN, Sezione di Torino, \phantom{more space} \\
Via P. Giuria, 1, I--10125 Torino, Italy}
\

\begin{abstract}

I discuss soft-gluon resummation and power corrections for event shape
distributions, mostly in $e^+ e^-$ annihilation. I consider
specifically the thrust, the $C$ parameter, and the class of
angularities, and show how factorization techniques and dressed gluon
exponentiation lead to predictive models of power corrections that are
firmly grounded in perturbative QCD. The scaling rule for the shape
function for angularities is derived as an example. Finally, I make a
few remarks on possible generalizations to hadron collisions, and on
their relevance to LHC studies.

\end{abstract}


\maketitle

\thispagestyle{fancy}


\section{INTRODUCTION}
\label{intro}

Quantum Chromodynamics is made special among phenomenologically
relevant field theories by the property of confinement. Given that the
lagrangian is written in terms of fields that do not appear in the
construction of the true asymptotic states of the theory, it may seem
surprising that perturbative calculations performed around the trivial
vacuum have any relevance at all. The predictive power of perturbative
QCD, in the presence of a kinematic scale $Q^2$ much larger than the
confinement scale $\Lambda^2$, is rescued by asymptotic freedom,
combined with quantum-mechanical incoherence and gauge invariance.
These are the necessary ingredients entering the proof of
factorization theorems~\cite{Collins:1989gx}, which are the
cornerstones of all PQCD calculations.

Factorization theorems in essence provide a bound for the parametric
size of nonperturbative corrections to high energy inclusive cross
sections. Such corrections are typically suppressed by powers of the
hard scale $Q^2$. It should be emphasized, however, that factorization
theorems are proven perturbatively, by examining the all-order
structure of long-distance singularities in Feynman diagrams. Their
phenomenological relevance must then rely upon the additional (if
plausible) assumption that confinement be a relatively soft process,
happening without a violent rearrengement of momentum configurations,
as colored particles evolve away from the hard scattering event.
Decades of experience with QCD phenomenology have taught us that this
assumption is very well borne out by the data. Indeed, granted this
assumption, the bound on the size of nonperturbative effects provided
by the factorization theorem is the first and simplest case of
nonperturbative information extracted form QCD by purely perturbative
methods.

The general idea that perturbation theory, whenever genuine all-order
information is available, can provide important clues to understand
nonperturbative effects, has subsequently been applied successfully in
a variety of situations. Typically, the perturbative expansion is
found to diverge, and the uncertainty in the physical prediction due
to this divergence is interpreted as a measure of the size of the
expected nonperturbative correction. Specifically, the nonperturbative
contribution must be ambiguous by an amount matching the uncertainty
in the perturbative prediction. The assumption that the actual size of
the nonperturbative corrections should be well represented by this
ambiguity is sometimes referred to as ultraviolet dominance of power
corrections~\cite{Beneke:1997sr}.

The all-order perturbative information required to begin any study of
power corrections has mostly been provided by two complementary
sources: renormalon-type calculations (reviewed in
\cite{Beneke:1998ui}), which roughly speaking target running-coupling
effects by summing up fermion bubble corrections to single gluon
emission, and soft gluon resummations, which make use factorization
and universality to compute leading multigluon contributions in the
soft and collinear limits (for a recent review, see
\cite{Laenen:2004pm}). Recently, it was shown that the two appoaches
can be combined \cite{Gardi:2001di}, yielding a strongly constrained
and rather elegant model of power corrections in the Sudakov region.

Event shape distributions in hard collisions are an especially
interesting class of observables for power correction studies, and
indeed a lot work has been done in the past several years on the
subject, especially in the context of $e^+ e^-$ annihilation and
DIS~\cite{Dasgupta:2003iq,Magnea:2002xt}. Event shape distributions,
in fact, provide a continuous interpolation between processes
featuring mostly hard, perturbative radiation and configurations
dominated by soft and collinear gluon emission. The corresponding
theoretical prediction must then be constructed matching a variety of
tools: NLO perturbative results for hard emissions, soft gluon
resummations when the value of the event shape forces radiation to be
soft, and finally models of power corrections very close to threshold.

In general, models of power corrections involve nonperturbative
parameters or functions, which must be determined from experiment,
much as one does with parton distributions. The predictive power of
these models must then rely on a degree of universality of soft
radiation, which is well understood in perturbation theory, and must
be assumend to hold nonperturbatively as well. By comparing theoretical
predictions for different but related event shapes one can then test
our understanding of QCD at or beyond the strict limits of
applicability of perturbation theory. 

The application of these techniques has lead in recent years to
quantitatively testable and quite successful models of power
corrections.  Here I will mostly discuss results obtained by Dressed
Gluon Exponentiation (DGE) \cite{Gardi:2001di}, as applied to thrust
\cite{Gardi:2001ny,Gardi:2002bg}, the C-parameter \cite{Gardi:2003iv},
and the class of angularities
\cite{Berger:2003pk,Berger:2004xf}. These examples show that current
tools lead to simple, analytical, quantitative results that can
readily be compared with experimental data. Most strikingly, leading
power corrections to angularity distributions obey a simple scaling
rule as a function of a continuous parameter, which gives a powerful
test of our understanding of soft QCD in electron-positron
annihilation. In Sect.~(\ref{soft}) I will briefly summarize the
formalism of shape functions for event shape distributions, and show
how DGE provides a renormalon model for shape functions incorporating
the constraints of NLL soft gluon resummation. I will use mostly
thrust as a working example, comparing at the end with similar results
obtained for the $C$ parameter. In Sect.~(\ref{angu}) I will discuss
the class of angularities and derive the scaling rule, and finally in
Sect.~(\ref{hadro}) I will briefly comment on possible extensions of
these techniques to hadron-hadron collisions.

\section{SOFT GLUON EFFECTS FOR EVENT SHAPE DISTRIBUTIONS}
\label{soft}

An event shape distribution is a weighted cross section, assigning a
prescribed value to a specific infrared and collinear safe combination
of the momenta of final state particles in a high energy collision. In
the case of $e^+ e^-$ annihilation, let $F_m (p_1, \ldots, p_m)$ be
one such combination, computed for an $m$-parton final state. The
distribution of the associated event shape $f$ is then
\beq
\frac{d \sigma}{d f} = \frac{1}{2 Q^2} \sum_m \int d {\rm LIPS}_m~
\overline{\left| {\cal M}_m \right|^2} ~\delta \left(f - F_m (p_1,
\ldots, p_m) \right)~,
\label{diste}
\eeq
where ${\cal M}_m$ is the appropriate matrix element.
In the following, I will consider event shapes $f$ which vanish in the
limit of a pencil-like two-jet event. A prime and well-known example
is $\tau = 1 - T$, with $T$ the thrust,
\beq
T = \max_{\hat{n}} \left[\frac{\sum_i \left\vert \vec{p}_i \cdot \hat{n}
      \right\vert}{Q} \right]~,
\label{thrust}
\eeq
which I will use below to illustrate the general features of the
approach.  Other shapes I will consider include the $C$-parameter,
\beq
C = 3 - \frac{3}{2} \sum_{i, j}
      \frac{(p_i \cdot p_j)^2}{(p_i \cdot Q) \, ( p_j \cdot Q)}~,
\label{cpar}
\eeq
which does not require a maximization procedure, and the one-parameter
class of angularities,
\beq
\tau_a = \frac{1}{Q} \sum_i (p_\perp)_i
       {\rm e}^{- |\eta_i| (1 - a)}~.
\label{classang}
\eeq
where rapidity $\eta_i$ and transverse momentum $p_{\perp,i}$ are
computed with respect to the thrust axis, so that for $a = 0$ one
verifies that $\tau_0 = \tau$.

The common feature of these event shapes, which opens the way to an
all-order perturbative analysis and to studies of power corrections,
is the fact that for small values of $f$ all radiation is constrained
to be soft or collinear. As a consequence, the distributions develop
double logarithmic singularities of Sudakov type, which can (and must)
be resummed thanks to the universal properties of soft radiation and
to the factorizability of the cross section in the Sudakov limit.  In
QCD, resummation displays the ambiguity of perturbation theory,
originating from the presence of the running coupling evaluated at
soft scales. This leads naturally to models of power corrections. I
will now illustrate the general features of the method using thrust as
an example.

\subsection{Resummation}

In order to resum singular contributions to the thrust distribution in
the limit $\tau \to 0$, one needs to take a Laplace transform, which
factorizes the $\delta$-function constraint fixing the value of
$\tau$. Logarithmic contributions then exponentiate according to
\beq
  \hspace{-1cm} \int_0^\infty d \, \tau {\rm e}^{- \nu \tau} \frac{1}{\sigma}
  \frac{d \sigma}{d \tau} \, = \, \exp \Bigg[ \int_0^1 \frac{d u}{u} \left( 
  {\rm e}^{- u \nu} - 1 \right) \Bigg( B \left(\alpha_s \left(u Q^2 \right) 
  \right) \, + \, 2 \, \int_{u^2 Q^2}^{u Q^2} \frac{d q^2}{q^2} \, A 
  \left(\alpha_s(q^2) \right) \Bigg) \Bigg]~.
\label{genexp}
\eeq
The pattern of exponentiation is highly nontrivial, since the double
logarithms of the ordinary perturbative expansion turn into single
logarithms in the exponent. Generically one finds a structure of the
form~\cite{Catani:1992ua}
\beq
\sum_k \alpha_s^k \sum_p^{2 k} c_{k p} L^p
       \rightarrow \exp \Big[ L \, g_1 (\alpha_s L) + g_2 (\alpha_s L) +
       \alpha_s g_3 (\alpha_s L) + \ldots \Big]~,
\label{pattern}
\eeq 
where $L$ is the logarithm of the transformed variable, $L = \log \nu$
in this case. Leading logarithms (LL) are generated to all orders by
the function $g_1$, which is completely determined by the knowledge of
the anomalous dimension $A(\alpha_s)$ to one loop. Next-to-leading
logarithms (NLL), corresponding to the function $g_2$, require the
knowledge of $A(\alpha_s)$ to two loops and $B(\alpha_s)$ to one
loop. NLL accuracy is the common standard for resummation of event
shape distributions. Note however that Sudakov resummation, expressed
here by \eq{genexp}, in general involves one more function, $D
\left(\alpha_s(u^2 Q^2)\right)$. This function is associated with
wide-angle soft gluon emission, and is process-dependent, unlike the
anomalous dimension $A$. To any finite logarithmic accuracy, the
contributions of $D$ can be reproduced by modifying $B$ in a
process-dependent manner, and for the event shapes discussed here $D$
does not give any contribution at NLL level. The fact that $D$ dpends
on the scale $u Q$, however, has important consequences on power
corrections, as discussed below.

\subsection{Power Corrections}

One can deduce from \eq{genexp} (where the integration variable $u$ in
the exponent plays the role of $\tau$) that for small values of $\tau$
there are two relevant momentum scales: $\tau Q^2$ and $\tau^2
Q^2$. This can be understood from the physical picture underlying
Sudakov factorization: at small $\tau$ gluon radiation can be
organized into jets of particles collinear to the primary partons,
with invariant mass proportional to $\sqrt{\tau} Q$, plus the
contribution of wide-angle soft gluons, characterized by their total
energy $\tau Q$. It is natural to expect that power corrections will
be organized by these two scales, and thus be of the form
$(\Lambda^2/(\tau Q^2))^m$ and $(\Lambda^2/(\tau^2 Q^2))^n$
respectively.  Clearly, when $\tau \sim \Lambda/Q$ all power
corrections of this second kind become important, and must be
collectively taken into account. Power corrections in the larger
scale, on the other hand, become important only when $\tau \sim
\Lambda^2/Q^2$, a value which is too small to be relevant for LEP
fits. The need for power corrections is apparent in \eq{genexp}, since
the integrals over momentum scales are perturbatively ill-defined
because of the Landau singularity of the running coupling. An elegant
way to summarize the nonperturbative information encoded in
\eq{genexp} was described in \cite{Korchemsky:1999kt}. The basic
assumptions are the applicability of the factorization underlying
\eq{genexp} all the way down to values of $\tau$ such that $\Lambda^2
\sim \tau^2 Q^2 \ll \tau Q^2 \ll Q^2$, and the existence of a
nonperturbative definition of the running coupling rendering the scale
integrals well defined. Consider then, for example, the term in \eq{genexp}
containing the anomalous dimension $A$. In order to disentangle
perturbative and nonperturbative domains, one can simply introduce a
factorization scale $\mu$, switch the order of the $q^2$ and $u$
integrations, and define
\beqa
  S (\nu,Q^2) & \equiv  & \int_0^1 \frac{d u}{u} \left( {\rm e}^{- u \nu} - 
  1 \right) \int_{u^2 Q^2}^{u Q^2} \frac{d q^2}{q^2} \,
  A \left( \alpha_s (q^2) \right) \, = \, 
  S_{\rm NP}(\nu/Q,\mu) \, S_{\rm PT}(\nu, Q,\mu)~, \nonumber \\ 
  S_{\rm NP}(\nu/Q,\mu) & \equiv & \int_0^{\mu^2} \frac{d q^2}{q^2}
  A \left( \alpha_s (q^2) \right) \int_{q^2/Q^2}^{q/Q}
  \frac{d u}{u} \left({\rm e}^{- u \nu} - 1 \right)
  \, = \, \sum_{n = 1}^\infty \frac{1}{n!} \left( \frac{\nu}{Q}
  \right)^n \lambda_n(\mu^2) \, + \, {\cal O} \left( \frac{\nu}{Q^2}
  \right)~.
\label{sudpar}
\eeqa
The last equality expresses a set of nonperturbative contributions
to the Sudakov exponent in terms of moments of the anomalous dimension
$A$ at low scales. These moments,
\beq      
  \lambda_n(\mu^2) = \frac{1}{n} \int_0^{\mu^2} d q^2 \, q^{n - 2}
  A \left( \alpha_s (q^2) \right)~,
\label{lamb}
\eeq
are not computable in perturbation theory: much like parton
distributions, they should be measured for a given observable at a
given factorization scale, and then used to predict different
observables, based on their universality properties. In general, the
full set of leading nonperturbative corrections will involve also
moments of the function $D$, which also parametrize power corrections
of the form $(\Lambda \nu/Q)^n$. These corrections thus have a
universal component, expressed in terms of the anomalous dimension
$A$, and a process-dependent component given by the function $D$. At
power accuracy, it is natural to disentangle the contributions of $B$
and $D$ by requiring that the function $B$ should be the same
appearing in the resummation formula for DIS structure functions,
where the corresponding $D$ function is known to
vanish~\cite{Gardi:2006jc}.  Expressions like \eq{sudpar} provide a
framework to test universality, or to construct specific models of
power corrections. To summarize the effects of the parameters
$\lambda_n (\mu^2)$ one can use them to build up a ``shape function'',
according to
\beq
  \exp \Big[ S_{\rm NP}(\nu/Q,\mu) \Big] \, \equiv \,
  \int_0^\infty d \epsilon \, {\rm e}^{- \nu \epsilon/Q} \,
  f_{\tau, {\rm NP}}(\epsilon, \mu)~.
\label{shpf}
\eeq 
Here $\e$ can be interpreted as the total energy carried into the
final state by soft gluons at scales below $\mu$. Confining oneself to
the leading power correction, corresponding to the first moment
$\lambda_1 (\mu^2)$, one recovers the result of the ``tube model''
\cite{Webber:1994zd}: that nonperturbative effects shift the
distribution away from the small $\tau$ region by an amount
proportional to the average energy carried away by soft
radiation. Subleading moments provide additional smearing.

\subsection{Dressed gluon exponentiation}

The shape function idea is very general, and can be used both to test
universality, by connecting power corrections to related event
shapes~\cite{Korchemsky:2000kp}, or to construct models based on
factorization and Lorentz invariance in specific
cases~\cite{Belitsky:2001ij}. One can get more detailed predictions by
making stronger assumptions: for example, one can apply a renormalon
model, and study the corresponding power corrections in the Sudakov
region. This is the basic idea underlying dressed gluon exponentiation
(DGE)~\cite{Gardi:2001di}. I will now summarize the basic steps of
this method, using thrust as an example.

First of all, one computes the single gluon contribution to the event
shape under study, for a gluon of nonvanishing virtuality $\xi =
k^2/Q^2$. This is the characteristic function of the dispersive
approach~\cite{Ball:1995ni,Dokshitzer:1995qm} to power corrections.
Since we are interested in the Sudakov region, we need to retain only
terms that are singular as $\tau \to 0$. Given the characteristic
function ${\cal F}(\xi, \tau)$, one can write a clean representation
of the single gluon cross section by introducing a Borel
representation for the strong coupling. One defines
\beq
\bar{A}(\xi Q^2) = \int_0^{\infty} d u \, \xi^{- u} \, \left( Q^2/\Lambda^2 
\right)^{- u} \, \frac{\sin \pi u}{\pi u} \, {\rm e}^{\kappa u} \,.
\label{axiq}
\eeq
This amounts to an analytic continuation of the strong coupling at the
scale $k^2$ from the euclidean to the timelike region, formally valid
in the large-$n_f$ limit. The factor ${\rm e}^{\kappa u}$ is
renormalization-scheme dependent, with $\kappa = 5/3$ in the $\MS$
scheme.  The single dressed gluon cross section is then
\beq
\frac{1}{\sigma} \frac{d \sigma}{d \tau} (\tau, Q^2) = - \frac{C_F}{2
\beta_0} \, \int_{0}^{1}{d \xi} \, \frac{d {\cal F}(\xi, \tau)}{d \xi}
\, \bar{A} (\xi Q^2) \, \equiv \, \frac{C_F}{2 \beta_0} \, \int_0^{\infty} 
d u \, \left( Q^2/\Lambda^2 \right)^{- u} \, B(\tau, u)\,,
\label{sdgf}
\eeq
where in the last equality I introduced the Borel function $B(\tau,
u)$, which contains the physical information on the thrust
distribution. The strategy of performing all integrals except the one
over the Borel parameter yields at the end a trasparent representation
for power corrections. In the case of thrust, the terms responsible
for logarithmic enhancements in $\dot{\cal F}(\xi, \tau)$ are given by
\beq
\left. \dot{\cal F}(\xi, \tau) \right|_{\log} =
2 \, \left(\frac{2}{\tau} - \frac{\xi}{\tau^2} - 
\frac{\xi^2}{\tau^3} \right)~,
\label{hjm1}
\eeq
which gives a Borel function of the form
\beq
\left. B(\tau, u) \right\vert_{\rm log} \, = \, 2 \, {\rm e}^{\kappa u} \,
  \frac{\sin \pi u}{\pi u} \, \left[ \frac{2}{u} \, \tau^{- 1 - 2 u} -
  \tau^{- 1 - u} \left(\frac2u + \frac1{1 - u} +
  \frac{1}{2 - u} \right) \right]\,.
\label{Bt}
\eeq
Note that already at this stage one can make several useful
observations.  Poles in $B(\tau, u)$ at, say, $u = u_0$, would
correspond to renormalon singularities in the distribution, and
expected power corrections of size $(\Lambda/Q)^{2 u_0}$; in fact,
$B(\tau, u)$ has no such poles: the would-be singularities at $u =
1,2$ are cancelled by the factor $\sin \pi u$, and poles at $u = 0$
cancel between the two terms in square brackets, because of the
infrared safety of thrust. Renormalons arise when taking moments of
the distribution, because of the convergence constraints on the Borel
integral at small values of $\tau$. One also notes that the first term
in \eq{Bt} is associated with soft wide-angle radiation, since it
generates in \eq{sdgf} terms proportional to $(\Lambda/(\tau Q))^{2
u}$.  Similarly, the second term in \eq{Bt} is associated with the
jet function, contaning collinear as well as soft enhancements.

The key step in DGE is to note that at LL level Sudakov resummation
yields a simple exponentiation of the one-gluon emission cross section
in moment space. One can then retain all large-$n_f$ information, and
the corresponding model of power corrections, in the Sudakov exponent
by simply using the single dressed gluon cross section as kernel of
exponentiation. One defines
\beq
\left( \frac{1}{\sigma} \frac{d \sigma}{d \tau}
  \right)_{\rm DGE} = \int_{ k - {\rm i} \infty}^{k + {\rm i} \infty}
  \frac{d \nu}{2 \pi {\rm i}} \, {\rm e}^{\nu \tau} \,
  \exp \left[ - E (\nu, Q^2) \right]\,,
\label{tau_DGE}
\eeq
where the Sudakov exponent is now given by
\beq
E (\nu, Q^2) \, = \, \int_0^{\infty} \, d \tau \,
\left( 1 - {\rm e}^{- \nu \tau} \right)
\left( \frac{1}{\sigma} \frac{d \sigma}{d \tau} \right)_{\rm SDG} 
\equiv \frac{C_F}{2 \beta_0} \, \int_0^{\infty} d u \,
\left(Q^2/\Lambda^2\right)^{- u} \, B_\tau (\nu, u) \, .
\label{S}
\eeq
Here the single dressed gluon cross section is defined by \eq{sdgf},
virtual corrections have been taken into account by subtracting the
value of the Laplace transform at $\nu = 0$, and in the second
equality the Borel function $B_\tau (\nu, u)$ for the Sudakov exponent
has been defined. As usual, all integrals are performed except the one
on the Borel parameter $u$. Although formally similar to \eq{sdgf} for
the single dressed gluon cross section, \eq{S} has a much richer
physical content, displayed by the nontrivial renormalon structure of
the Borel function $B_\tau (\nu, u)$. This is a consequence of the
fact that exponentiation, subject to the constraint of energy
conservation, has promoted the single gluon result to a genuine
approximation for multigluon emission\footnote{In fact, the exponent
$E (\nu, Q^2)$ has a natural interpretation in terms of the Borel
representation of Sudakov anomalous dimensions associated respectively
with soft and collinear radiation~\cite{Gardi:2006jc}.  Corrections
subleading in $n_f$ change the the logarithmic behavior of the cross
section as well as the size of the residues of poles in the Borel
plane, but they are not expected to modify the analytic structure of
$B_\tau (\nu, u)$, which determines which power corrections are
actually present.}.  In the specific case of the thrust, the result
for the Borel function is
\beq
B_\tau (\nu, u) \, = \, 2 \, {\rm e}^{\kappa u} \, \frac{\sin \pi u}{\pi u}
\left[\Gamma(- 2 u) \left(\nu^{2 u} - 1 \right) \frac{2}{u}
- \Gamma(- u) \left(\nu^u - 1\right) \left(\frac2u + \frac1{1 - u}
+ \frac{1}{2 - u} \right) \right]\,.
\label{bnut}
\eeq
It is useful to compare this result with the one for the
C-parameter~\cite{Gardi:2003iv}
\beq
B_c (\nu, u) \, = \, 2 \, {\rm e}^{\kappa u} \,
  \frac{\sin \pi u}{\pi u} \, \left[\Gamma(- 2 u) \left(\nu^{2 u} - 1
  \right) 2^{1 - 2 u} \frac{\sqrt{\pi} \Gamma(u)}{\Gamma(\frac12 + u)}
  \, - \, \Gamma(- u) \left({\nu}^{u} - 1 \right)
  \left(\frac2u + \frac1{1 - u} + \frac{1}{2 - u} \right) \right]\,.
\label{bnuc}
\eeq
Clearly, Eqs.~(\ref{bnut}) and (\ref{bnuc}) are very similar; there
are, however, important differences, which highlight the degree of
universality to be expected in comparing the two distributions, and
which can in principle be tested experimentally. First of all, the
Borel functions contain perturbative information on Sudakov
logarithms: although the exponentiation was performed assuming
independent gluon radiation, it can be shown~\cite{Gardi:2001ny} that
one can upgrade the formalism to NLL accuracy by simply replacing the
running coupling \eq{axiq} with the two-loop expression, and by
changing renormalization scheme, including in the constant $\kappa$
the contribution of terms singular as $x \to 1$ in the NLO
Altarelli-Parisi splitting function (the ``gluon bremsstrahlung''
scheme~\cite{Catani:1990rr}). Beyond NLL, the coefficients of all
subleading logarithms can be computed in the large $n_f$ limit, by
simply expanding the Borel function in powers of $u$, and replacing
$u^n \to n! (b_0 \alpha_s/\pi)^{n + 1}$. Computing subleading
logarithms uncovers the factorial growth of their coefficients, and
can be used to gauge the reliability of perturbative resummation in
different kinematical regimes. Next, one may observe that the infrared
safety of $\tau$ and $C$ is once again reflected in the cancellation
of the poles of both Borel functions at $u = 0$. One also notes that
wide-angle soft radiation and collinear gluons contribute as before
two separate terms, and the jet functions (the terms proportional to
$\Gamma(- u)$) are identical for the two observables\footnote{Note
that $B_c (\nu, u)$ is computed for a rescaled $C$ parameter, $c =
C/6$}. They contribute renormalons at $u = 1,2$, corresponding to
exponentiated power corrections of the form $(\Lambda^2 \nu/
Q^2)^{1,2}$. Soft gluon contributions, on the other hand, have
renormalons at $u = m/2$, for all odd values of $m$, yielding the
leading power corrections $(\Lambda \nu/Q)^m$, and they are
quantitatively different for $\tau$ and $C$, distinguishing the two
observables. Specifically, by taking the ratio of the two ``soft
functions'' (the terms proportional to $\Gamma(- 2 u)$) and expanding
in powers of $u$, one can verify that the two observables begin to
differ perturbatively at NNLL level, as predicted
in~\cite{Catani:1998sf}, but the growth of the coefficients of further
subleading logarithms is weaker for the $C$-parameter than for the
thrust. Similarly, if one boldly takes the large-$n_f$ residues of the
poles of the Borel functions as a reasonable estimate of the size of
the corresponding power corrections, one observes that $(\Lambda
\nu/Q)^m$ corrections are systematically smaller for the $C$
parameter: the two shape functions should therefore differ, and one
expects that the resummed perturbative prediction, as well as the
approximation of the shape function by a constant shift, should work
better phenomenologically for $C$ than for $\tau$.

\section{THE CLASS OF ANGULARITIES}
\label{angu}

The discussion in Sect.~(\ref{soft}) illustrates the predictive power
of DGE. Another interesting application concerns angularities, whose
definition, \eq{classang}, can be rewritten for massless particles as
\beq
  \tau_a \, = \, \frac{1}{Q} \, \sum_i \omega_i \left( \sin
  \theta_i \right)^a \left( 1 - \left| \cos \theta_i \right|
  \right)^{1 - a}~,
\label{barfdef}
\eeq
with $\theta_i$ the angle with respect to the thrust axis.
Angularities (so christened in \cite{Berger:2004xf}) were introduced
in \cite{Berger:2003iw} as auxiliary shape variables used to tame
nonglobal logarithms~\cite{Dasgupta:2001sh} for observables related to
out-of-jet energy flow. They have several remarkable features, which
make them very interesting for our understanding of QCD at the edge of
the perturbative domain . First of all, they are characterized by a
tunable parameter $a$, which can be used to interpolate between
different shapes, or to bring to focus specific momentum
configurations in a continuous way. The parameter $a$ must satisfy $a
< 2$ for infrared safety, and a tighter restriction $a <1$ is required
in order to preserve a relatively simple resummation in the Sudakov
region, in the form of \eq{genexp}: for $1 \leq a < 2$ further
logarithmic singularities associated with jet recoil must be taken
into account~\cite{Dokshitzer:1998kz}. For $a = 1$, one recognizes
that $\tau_1 = B$, the broadening; for $a = 0$, $\tau_0 = 1 - T$; for
negative $a$, events dominated by high rapidity give increasingly
suppressed contributions to $\tau_a$, which in turn suppresses power
corrections of collinear origin; finally, for $a \to - \infty$ the
distribution becomes a $\delta$ function at $\tau_a = 0$, with a
strenght given by the total cross section.

Remarkably, although the relative weights of rapidity and transverse
momentum change with $a$, it is possible to derive a resummation
formula~\cite{Berger:2003iw} of the form of \eq{genexp}, valid for $a
< 1$.  Indeed, at NLL accuracy one can write
\beq
  \ln \big[ \tilde{\sigma} \left(\nu, a \right) \big] \, = \,
  \int\limits_0^1 \frac{d u}{u} \, \Bigg[ \,
  B \left(\as(u Q^2)\right) \left( {\rm e}^{- u \, \nu^{2/(2 - a)} } -1 \right)
  \, + \, 2 \, \int_{u^2 Q^2}^{u Q^2} \frac{d q^2}{q^2} \,
  A \left(\as (q^2) \right) \left( {\rm e}^{- u^{1 - a} \nu 
  \left( q/Q \right)^a } - 1 \right) \Bigg]~.
\label{thrustcomp}
\eeq
The $a$ dependence of Sudakov logarithms is clearly nontrivial: as an
example, the function $g_1 (\alpha_s L)$ responsible for leading
logarithms in \eq{pattern} is given by
\beq
g_1 ( x, a ) \, = \, - \, \frac{4}{\beta_0} \, \frac{2 - a}{1 - a} \,
  \frac{A^{(1)}}{x} \, \Bigg[ \frac{1 - x}{2 - a} \,
  \ln \left( 1 - x \right) \, - \, \left(1 - \frac{x}{2 - a}\right) 
  \ln \left(1 - \frac{x}{2 - a}\right) \Bigg] \, .
\label{g1a}
\eeq
Power corrections, however, turn out to have a much simpler $a$
dependence~\cite{Berger:2003pk}.  Performing the analysis leading to
\eq{lamb}, one easily finds that in the nonperturbative region all
moments of the anomalous dimension $A$ are multiplied by a simple
common factor
\beq      
  \lambda_n^{(a)} (\mu^2) \, = \, \frac{1}{1 - a} \,\frac{1}{n} \, 
  \int_0^{\mu^2} d q^2 \, q^{n - 2} A \left( \alpha_s (q^2) \right) \,
  = \,  \frac{1}{1 - a} \, \lambda_n^{(0)} (\mu^2) ~.
\label{lamba}
\eeq 
As a consequence, the Laplace transform of the shape function defined 
by \eq{shpf} obeys a simple and remarkable scaling rule
\beq
  \tilde{f}_{a, {\rm NP}} \left(\frac{\nu}{Q}, \kappa\right) = \left[
  \tilde{f}_{0, {\rm NP}} \left(\frac{\nu}{Q}, \kappa\right)
  \right]^{1/(1 - a)}~,
\label{rule}
\eeq
which should be experimentally testable without great effort using
existing LEP data. Lacking a direct experimental analysis, the scaling
rule was tested against the output of {\tt PYTHIA}, with positive
results~\cite{Berger:2003pk,Berger:2003gr}. The physical picture
underlying the scaling rule is appealing, and once again reminiscent
of the ``tube model''. The relevant feature of the radiation pattern
is boost invariance, which applies to soft gluons emitted in the
two-jet limit, since such emissions are correctly represented by the
eikonal approximation. This means that soft gluons contribute to the
event shape a rapidity-independent amount, and in turn the integration
over rapidity simply measures the size of the region where gluons can
be emitted without strongly affecting the event shape. This region
scales with $(1 - a)^{-1}$. The derivation of the scaling rule
neglects correlations between gluons emitted into the opposite
hemispheres defined by the thrust axis, which however are expected to
become important only for $a \geq 1$, the region excluded by the
present treatment. The main assumption is then that nonperturbative
soft radiation should share the property of boost inveriance with the
relatively harder perturbative component which is treated by
resummation. This is the nonperturbative property that would be
directly tested by an experimental study of \eq{rule}.

One can go further, and study subleading power corrections, mostly
related to radiation collinear to the hard jets. Applying the same
method to the anomalous dimension $B$, and to the subleading terms
generated by $A$, once again one finds a simple pattern: the Laplace
transform of the cross section can be expressed in factorized form,
introducing a subleading shape function, as
\beq
  \tilde{\sigma} \left(\nu, a\right) = \tilde{\sigma}_{\rm PT}
  \left(\nu, \kappa, a\right) \, \tilde{f}_{a, {\rm NP}}
  \left(\frac{\nu}{Q},\kappa\right) \, \tilde{g}_{a, {\rm NP}}
  \left(\frac{\nu}{Q^{2 - a}},\kappa\right)~.
\label{sub}
\eeq
Clearly, as $a$ becomes large and negative, collinear power
corrections are expected to become more and more negligible, and the
scaling rule \eq{rule} is expected to hold with increasing accuracy.

In the light of the discussion of Sect.~(\ref{soft}), it is
interesting to verify whether these nice features of angularities are
preserved in specific models for the shape function, such as DGE.
This is particularly relevant in this case, since the key result,
\eq{rule}, is related to boost invariance, which is broken by the
formal introduction of gluon virtuality, which is a necessary tool of
renormalon analysis. The study of angularities with DGE was performed
in~\cite{Berger:2004xf}. It is technically nontrivial, since the
interplay of the parameter $a$ with gluon virtuality and phase space
constraints makes it difficult to extract the $a$ dependence
analytically. The first step is to find an appropriate generalization
of the definition of angularity to the case of single massive gluon
emission.  Such definition should have the correct limit as $\xi \to
0$, reduce to known results for thrust as $a \to 0$, and be simple
enough to keep the computational task manageable. At one loop, one
such definition is
\beq
    \tau_a = \frac{(1 - x_i)^{1 - a/2}}{x_i}
    \left[(1 - x_j - \xi)^{1 - a/2} (1 - x_k + \xi)^{a/2} + 
    (j \leftrightarrow k) \right]~,
\label{defo}
\eeq
where $x_n = 2 p_n \cdot Q/Q^2$ are the customary energy fraction
variables, and the definition applies to the phase space region where
the gluon is soft; $x_i$ is then the (anti)quark energy fraction.
With this definition, it is possible to construct the Borel function
for the Sudakov exponent for angularities, in analogy with
Eqs.~(\ref{bnut}) and (\ref{bnuc}). Remarkably, the soft component of
the Borel function, responsible for all leading power corrections, is
just the expected rescaling of \eq{bnut},
\beq
    B_a^{\rm soft} (\nu, u) = \frac{1}{1 - a} \left[ 2 \,
    {\rm e}^{\kappa u} \, \frac{\sin \pi u}{\pi u} \, \Gamma(- 2 u)
    \left(\nu^{2 u} - 1 \right) \frac{2}{u} \right]~,
\label{scale}
\eeq
which leads once again to the scaling rule, \eq{rule}. Collinear power
corrections are much more difficult to handle analitycally, and the
collinear counterpart of \eq{scale} can at best be expressed in terms
of a one-dimensional integral representation, which reduces to
combinations of hypergeometric functions for rational values of
$a$. This is however enough to classify the singularities in the Borel
parameter $u$, and thus the pattern of power corrections. One indeed
finds that all these subleading power corrections can be organized in
a single shape function $\tilde{g}$, depending only on the combination
$\nu/Q^{2 - a}$, as in \eq{sub}. DGE thus confirms the general scaling
behavior found from resummation, showing that the introduction of
gluon virtuality does not spoil the effects of boost invariance in the
Sudakov limit. Thrust, jet masses, angularities and the $C$ parameter
are all found to have closely related pattern of power corrections,
highlighted by the scaling rule relating generic angularities to the
thrust. Clearly, this is a highly predictive framework, and our
understanding of soft radiation in the two-jet limit can be put to
stringent tests.

\section{HADRON COLLIDER EVENT SHAPES}
\label{hadro}

As we approach the expected date for the start up of the Large Hadron
Collider at CERN, it is natural and appropriate to ask whether tools
like those described here could be applicable in the environment of
hadron collisions, and, if so, to what extent power corrections might
be relevant to our understanding of the data, at the extreme energies
available at the LHC. Beginning with the second question, one might
naively observe that $\Lambda/Q$ must be a very small number for any
reasonable value of the hard scale $Q$ that one might envisage at the
LHC. It would not be wise, however, to neglect power correction
studies on this basis, for at least three different reasons.

First of all, it has already been shown at the Tevatron that power
corrections have an impact even on observables which are largely
dominated by high $p_\perp$ events, and even at very high
energy~\cite{Mangano:1999sz}. In Ref.~\cite{Mangano:1999sz}, Mangano
considered the single-jet inclusive $E_\perp$ distribution, comparing
data at different CM energies. One sees that ratios of cross sections
at different energies do not have the proper scaling behavior dictated
by NLO QCD, but the correct behavior can be recovered by including a
power correction determined by a single parameter associated with the
normalization of the jet transverse energy. The reason is that even a
small shift in the jet $E_\perp$ is amplified in the distribution by
the fact that the cross section is falling steeply for increasing
$E_\perp$, so that $\delta \sigma/\sigma \sim - n \, \delta E_\perp$
if $\sigma \sim E_\perp^{-n}$. One can expect that in general power
corrections will be important for the determination of jet energy
scales, and in turn accurate knowledge of these scales may well turn
out to be crucial for many high energy studies.

The second reason is that almost any LHC observable will require,
before it can be compared with a theoretical prediction, a subtraction
of all hadronic activity unrelated to the hard scattering, loosely
referred to as ``underlying event''. There is currently very little,
if any, theoretical control on the underlying event, which at the LHC
will contain a mix of multiple parton scattering, beam-beam
interactions and soft radiation associated with the selected hard
process. Even Monte-Carlo methods have difficulties in finding the
proper tuning to describe this kind of physics (see, for
example,~\cite{Alekhin:2005dx}). In this context, the lesson of power
correction studies in the gentler environment of $e^+ e^-$
annihilation or DIS is that we might learn to discriminate between the
different components of soft radiation in hadron collisions. On the
one hand there are soft and collinear gluons associated with the hard
scattering event: their effects are in principle computable in PQCD,
using generalizations of the known techniques, and their distribution
in phase space and in the space of color configurations will be
nontrivial and predictable. On the other hand, there is soft radiation
which is in practice out of reach for the techniques of PQCD, such as
minijets due to multiple parton scattering or soft gluons arising from
beam-beam interactions. This second kind of radiation fills phase
space with a high degree of uniformity, and will have to be modelled
with different techniques, including Monte-Carlo tools. This kind of
statistical modelling is bound to be more successful if we can first
achieve a better understanding of the ``pure'' hard scattering
process, including the energy and color flow that it generates at all
scales.

Finally, it should be emphasized that the physics of event shapes at
hadron colliders is interesting for its own sake, as a probe to
understand hadronization in terms of both momentum and color flow. It
is well known that for hadron collisions, where most processes involve
four partons already at Born level, Sudakov resummation is expressed
in terms of anomalous dimension matrices that tie together color
exchange and momentum flow. This interplay is bound to influence the
pattern of power corrections as well, and studies of this kind may
well lead to deep insights into the mechanics of color neutralization
and hadron formation.

Preliminary studies of resummed event shapes at hadron colliders have
already been performed~\cite{Banfi:2004nk,Zanderighi:2006mx}, and a
generalization of the notion of angularity to a hadronic environment
has been proposed~\cite{Berger:2005ct}. For most of these proposals, a
primary concern is that of suppressing the contributions of particles
close to the beam axis, where beam remnants interfere with all
measurements. It should be noted, however, that soft radiation
associated with the underlying event tends to fill phase space,
including regions separated in rapidity from both the beam and the
high-$p_\perp$ jets.  It would therefore be of great interest to find
event shapes designed to focus on this kind of wide-angle radiation,
where it would be most useful to disentangle soft gluons generated by
the hard scattering from the genuine underlying event (a pioneering
study with a similar goal is~\cite{Marchesini:1988hj}). A promising
avenue of investigation might be the use of observables such as those
introduced in~\cite{Berger:2003iw}, joint distributions correlating
energy flow in a chosen angular region $\Omega$ with a standard event
shape such as ordinary angularity. The form of such a correlation is
\beq
  \sigma \left( \epsilon_1, \epsilon_2, a \right) = \frac{1}{2 s} \sum_N
  \overline{\left| M(N) \right|^2} \,\,
  \delta (\epsilon_1 - f_\Omega (N)) \,\,  \delta(\epsilon_2 -
  \tau_a^{(1)} (N) - \tau_a^{(2)} (N))~,
\label{esefc}
\eeq
where $f_\Omega (N)$ is an observable related to energy flow into the
angular region $\Omega$, away from hard jets, while $\tau_a^{(i)} (N)$
are the contributions to angularity from the two hemispheres defined
by the thrust axis.

In a hadronic environment, one might envisage measuring angularities
with respect to the current jet axis, as suggested
in~\cite{Berger:2005ct}, or even introducing a third parameter
$\epsilon_3$ to constrain radiation near the beam remnants. Tuning the
various parameters and energy threshold in such observables one would
be able to focus on different regions of phase space, forcing the
observable to be more or less inclusive for soft gluons, or for
particles collinear either to the beam or to the current jets.  Notice
that boost invariance, underlying the scaling rule for angularities,
is lost in hadron collisions, where correlations between beam jets and
current jets must be taken into account.

Clearly, event shape studies at hadron colliders are in their early
days.  I believe that such studies will be both instrumental to
further our understanding of QCD, and very helpful in order to exploit
the full potential of the LHC, a task which will require a solid
understanding of strong interactions as much as good skills in the
building and testing of new physics models.

\begin{acknowledgments}

\noindent I would like to thank Carola Berger and Einan Gardi, who
worked with me on the research described here, and went over the draft
of this contribution, spotting several misprints and one significant
inaccuracy, now fixed. I also thank George Sterman and Mrinal Dasgupta
for sharing their insights, and the organizers and FRIF for the
opportunity granted by the Workshop, which was conducted in a friendly
and constructive atmosphere. This work was supported in part by MIUR
under contract $2004021808\_009$.

\end{acknowledgments}


\begin{thebibliography}{99}

\bibitem{Collins:1989gx}
  J.~C.~Collins, D.~E.~Soper and G.~Sterman,
  {\it Adv. Ser. Direct. High Energy Phys.} {\bf 5} (1988) 1,
  {\tt hep-ph/0409313}.

\bibitem{Beneke:1997sr}
  M.~Beneke, V.M.~Braun and L.~Magnea,
  {\it Nucl. Phys.} {\bf B 497} (1997) 297, {\tt hep-ph/9701309}.

\bibitem{Beneke:1998ui}
  M.~Beneke,
  {\it Phys. Rept.} {\bf 317} (1999) 1, {\tt hep-ph/9807443}.

\bibitem{Laenen:2004pm}
  E.~Laenen,
  {\it Pramana} {\bf 63} (2004) 1225.

\bibitem{Gardi:2001di}
  E.~Gardi,
  {\it Nucl. Phys.} {\bf B 622} (2002) 365, {\tt hep-ph/0108222}.

\bibitem{Dasgupta:2003iq}
  M.~Dasgupta and G.P.~Salam,
  {\it J. Phys.} {\bf G 30} (2004) R143, {\tt hep-ph/0312283}.

\bibitem{Magnea:2002xt}
  L.~Magnea, in {\it Proceedings} of IFAE 2002, SIF,
  ed. M Cacciari {\it et. al}, p. 143,
  {\tt hep-ph/0211013}.

\bibitem{Gardi:2001ny}
  E.~Gardi and J.~Rathsman,
  {\it Nucl. Phys.} {\bf B 609} (2001) 123, {\tt hep-ph/0103217}.

\bibitem{Gardi:2002bg}
  E.~Gardi and J.~Rathsman,
  {\it Nucl. Phys.} {\bf B 638} (2002) 243, {\tt hep-ph/0201019}. 

\bibitem{Gardi:2003iv}
  E.~Gardi and L.~Magnea,
  {\it JHEP} {\bf 0308} (2003) 030, {\tt hep-ph/0306094}.

\bibitem{Berger:2003pk}
  C.F.~Berger and G.~Sterman,
  {\it JHEP} {\bf 0309} (2003) 058, {\tt hep-ph/0307394}.

\bibitem{Berger:2004xf}
  C.~F.~Berger and L.~Magnea,
  {\it Phys. Rev.} {\bf D 70} (2004) 094010, {\tt hep-ph/0407024}.

\bibitem{Catani:1992ua}
  S.~Catani, L.~Trentadue, G.~Turnock and B.R.~Webber,
  {\it Nucl. Phys.} {\bf B 407} (1993) 3.

\bibitem{Korchemsky:1999kt}
  G.P.~Korchemsky and G.~Sterman,
  {\it Nucl. Phys.} {\bf B 555} (1999) 335,
  {\tt hep-ph/9902341}.

\bibitem{Gardi:2006jc}
  E.~Gardi, these proceedings,
  {\tt hep-ph/0606080}.

\bibitem{Webber:1994zd}
  B.~R.~Webber,
  {\tt hep-ph/9411384}.

\bibitem{Korchemsky:2000kp}
  G.P.~Korchemsky and S.~Tafat,
  {\it JHEP} {\bf 0010} (2000) 010, {\tt hep-ph/0007005}.

\bibitem{Belitsky:2001ij}
  A.V.~Belitsky, G.P.~Korchemsky and G.~Sterman,
  {\it Phys. Lett.} {\bf B 515} (2001) 297, {\tt hep-ph/0106308}.

\bibitem{Ball:1995ni}
  P.~Ball, M.~Beneke and V.M.~Braun,
  {\it Nucl. Phys.} {\bf B 452} (1995) 563, {\tt hep-ph/9502300}.

\bibitem{Dokshitzer:1995qm}
  Y.L.~Dokshitzer, G.~Marchesini and B.R.~Webber,
  {\it Nucl. Phys.} {\bf B 469} (1996) 93, {\tt hep-ph/9512336}.

\bibitem{Catani:1990rr}
  S.~Catani, B.~R.~Webber and G.~Marchesini,
  {\it Nucl. Phys.} {\bf B 349} (1991) 635.

\bibitem{Catani:1998sf}
  S.~Catani and B.R.~Webber,
  {\it Phys. Lett.} {\bf B 427} (1998) 377, {\tt hep-ph/9801350}.

\bibitem{Berger:2003iw}
  C.F.~Berger, T.~Kucs and G.~Sterman,
  {\it Phys. Rev.} {\bf D 68} (2003) 014012, {\tt hep-ph/0303051}.

\bibitem{Dasgupta:2001sh}
  M.~Dasgupta and G.~P.~Salam,
  {\it Phys. Lett.} {\bf B 512} (2001) 323, {\tt hep-ph/0104277}.

\bibitem{Dokshitzer:1998kz}
  Y.~L.~Dokshitzer, A.~Lucenti, G.~Marchesini and G.~P.~Salam,
  {\it JHEP} {\bf 9801} (1998) 011, {\tt hep-ph/9801324}.

\bibitem{Berger:2003gr}
  C.F.~Berger and G.~Sterman,
  {\tt hep-ph/0310058}.

\bibitem{Mangano:1999sz}
  M.~L.~Mangano,
  {\tt hep-ph/9911256}.

\bibitem{Alekhin:2005dx}
  S.~Alekhin {\it et al.},
  {\tt hep-ph/0601012}.

\bibitem{Banfi:2004nk}
  A.~Banfi, G.~P.~Salam and G.~Zanderighi,
  {\it JHEP} {\bf 0408} (2004) 062, {\tt hep-ph/0407287}.

\bibitem{Zanderighi:2006mx}
  G.~Zanderighi, these proceedings,
  {\tt hep-ph/0605332}.

\bibitem{Berger:2005ct}
  C.~F.~Berger,
  {\it Mod. Phys. Lett.} {\bf A 20} (2005) 1187, {\tt hep-ph/0505037}.

\bibitem{Marchesini:1988hj}
  G.~Marchesini and B.~R.~Webber,
  {\it Phys. Rev.} {\bf D 38} (1988) 3419.

\end{thebibliography}
\end{document}